\documentclass[12pt,aps,prd,superscriptaddress,showpacs,longbibliography,floatfix,nofootinbib]{revtex4-1}

\usepackage[utf8]{inputenc}
\usepackage{slashed}
\pdfoutput=1

\usepackage{color}
\usepackage{graphicx}   
\usepackage{bm}
\usepackage{amsmath}
\usepackage{amsfonts}
\usepackage{eufrak}
\usepackage{hyperref}

\newcommand{\be}{\begin{equation}}
\newcommand{\ee}{\end{equation}}
\newcommand{\ba}{\begin{eqnarray}}
\newcommand{\ea}{\end{eqnarray}}

\begin{document}

\title{On the Gravitational Wave to Matter Coupling of Superfluid Fermi Gases Near Unitarity}

\author{Scott Lawrence}
\email{scott.lawrence-1@colorado.edu}
\affiliation{Department of Physics, University of Colorado, Boulder, Colorado 80309, USA}
\author{Paul Romatschke}
\email{paul.romatschke@colorado.edu}
\affiliation{Department of Physics, University of Colorado, Boulder, Colorado 80309, USA}
\affiliation{Center for Theory of Quantum Matter, University of Colorado, Boulder, Colorado 80309, USA}

\begin{abstract}
It is well known that gravitational waves distort equilibrium matter globally, making them amenable to detection with laser interferometers. Less well known is the fact that gravitational waves create local non-equilibrium stresses inside matter, which could conceivably lead to alternative detection methods. The gravitational wave to matter coupling $\kappa$ is a transport coefficient depending on the material, and is poorly known for most substances. In the present work, we calculate $\kappa$ for a superfluid Fermi gas near unitarity using large-$N$ techniques, finding $\kappa= \frac{n}{12m}$, with $n$ the number density and $m$ the mass of the fermion, matching the result for free Dirac fermions at zero temperature. Our prediction is amenable to non-perturbative theoretical as well as experimental tests.
\end{abstract}

\maketitle

\section{Introduction}

Detection of gravitational waves using laser interferometers is by now a mature field with observed binary black hole signals, black hole-neutron star signals and binary neutron star signals~\cite{LIGOScientific:2016aoc,LIGOScientific:2017ycc,LIGOScientific:2017zic,LIGOScientific:2020zkf}. In a historical context, however, this is an extremely recent development, given that gravitational waves were predicted by Einstein in 1916~\cite{Einstein:1916cc}, and that there were many unsuccessful attempts to detect gravitational waves for a century after the prediction was made.

Now that we have observational proof that gravitational waves do exist, it may be time to study their effects on physical systems in a more detailed manner, even though these effects will invariably be extremely weak.

The focus of this work is one such effect, the coupling of gravitational waves to matter. Similar to electromagnetic fields leading to effects in matter such as polarization, gravitational fields produce local stresses in an otherwise isotropic medium. Historically, this effect was first identified as a mathematical consequence of requiring consistent field equations for fluid dynamics \cite{Baier:2007ix,Bhattacharyya:2007vjd}, requiring a contribution to the stress-tensor
\be
\label{tij}
T^{ij}=\kappa \left[R^{\langle ij \rangle}-2 R^{t \langle ij \rangle t}\right]\,,
\ee
where the $R$'s denote the Ricci and Riemann curvature tensors, respectively, $\langle\rangle$ denotes symmetric traceless projection, $i=\left\{x,y,z\right\}$ and $t$ denotes the time-component\footnote{While (\ref{tij}) is required for consistency, this does not preclude the possibility of $\kappa=0$ for \textit{certain} material conditions.}. Because the curvature tensors are sensitive to the passing of a gravitational wave, (\ref{tij}) unequivocally implies that a medium will respond with a local non-vanishing stress $T^{ij}$ as long as $\kappa\neq 0$. For this reason, we refer to $\kappa$ as the (material-dependent) gravitational-wave to matter coupling.

Besides controlling the coupling of gravitational waves to matter, $\kappa$ possesses a dual role as a second-order transport coefficient. In general, transport coefficients control the real-time response of a system subject to a perturbation, with familiar examples being conductivities, diffusion coefficients and viscosities. However, these familiar examples encode only the linear (first-order) response of the system to a perturbation, whereas for real systems non-linear contributions (second-order, third-order, etc.) will also be present, and can be important when gradients are large. The gravitational-wave to matter coupling $\kappa$ is an example of a second-order transport coefficient, controlling the strength of the system's response to second order in a perturbation in flat space. At first glance, such a dual role of a transport coefficient for seemingly unrelated phenomena (coupling to gravitational waves and second-order flat-space perturbation) may seem strange, but we remind the reader of the more well-known example of the Einstein relation which also serves such a dual purpose (controlling both the diffusion coefficient and the conductivity). For the purpose of this work, the dual role of $\kappa$ can be exploited to easily obtain $\kappa$ by calculating flat-space correlation functions to second order in gradients.

A non-vanishing value for $\kappa$ was calculated for ${\cal N}=4$ Super-Yang-Mills at large 't Hooft coupling~\cite{Baier:2007ix,Finazzo:2014cna,Grozdanov:2014kva,Grieninger:2021rxd}, and more recently $\kappa=-\frac{13 N m_B^2}{2520 \pi^2}$ with $m_B$ the in-medium boson mass was found analytically for the large-$N$ limit of the interacting $O(N)$ model~\cite{Romatschke:2019gck}. As for theories that actually occur in nature, results for $\kappa$ have been reported for  Yang-Mills theory~\cite{Romatschke:2009ng,Philipsen:2013nea}, free bosons, and free Dirac fermions both at zero chemical potential \cite{Moore:2012tc,Kovtun:2018dvd} and zero temperature \cite{Shukla:2019shf}.

Curiously, $\kappa$ is \emph{not} known for non-relativistic two-component fermions near unitarity. Given the tremendous success of Fermi gas experiments in obtaining transport properties \cite{Papp_2008,Riedl_2008,Cao_2011,Gaebler_2010,Ku_2012,Patel_2020,Kuhn_2020,1912.06131,Del_Pace_2021,Navon_2021,Wang_2022}, and the potential application to using said experiments as a gravitational-wave detector, this provides motivation for calculating $\kappa$ for Fermi gases near unitarity.

Since no information about $\kappa$ is currently available for Fermi gases near unitarity, this study is exploratory in nature. For this reason, we find it acceptable to perform calculations for $\kappa$ in the R0 approximation\footnote{Also known as leading large $N$ or mean-field approximation~\cite{veillette2007large,nikolic2007renormalization}.} that was used to determine this quantity for the $O(N)$ model~\cite{Romatschke:2019rjk,Romatschke:2019gck}. The approximation scheme is systematically improvable in principle should one desire more accurate results for $\kappa$ in the future\footnote{In particular, the R4 resummation scheme employed~\cite{Romatschke:2021imm} will resum all contributions to $\kappa$ of order $N^{-1}$.}. As an alternative to  the analytic approach pursued in this work, we note that it would also be possible to extract $\kappa$ numerically by employing techniques suitable to obtain non-perturbative four-point correlation functions, similar to what has been done in lattice QCD~\cite{Philipsen:2013nea} and for shear viscosity in the unitary Fermi gas~\cite{Wlazlowski:2012jb}, see the discussion in section \ref{sec:discussion}.

Throughout this work we will be using natural units where $\hbar=c=k_B=1$, and convert to S.I.~units only when discussing applications of the results that have overlap with measurable quantities.


\section{Calculation}

Let us consider a Fermi gas in three spatial dimensions with the Hamiltonian density
\be
\label{hamiltonian}
   {\mathcal H}=\sum_{s=\uparrow,\downarrow}\psi^\dagger_s({\bf x}) \left(-\frac{\nabla^2}{2m}\right)\psi_s({\bf x})+
   \frac{4 \pi a_s}{m} \psi^\dagger_\uparrow ({\bf x})\psi^\dagger_\downarrow ({\bf x})\psi_\downarrow ({\bf x})\psi_\uparrow ({\bf x})\,,
\ee
where $\psi^\dagger_s,\psi_s$ are the fermionic creation and annihilation operators for spin (or hyperfine state) $s=\;\uparrow,\downarrow$, respectively, $m$ is the fermion mass, and $a_s<0$ is the s-wave scattering length. Being  field theorists, we much prefer discussing the properties of this system in terms of the grand-canonical partition function, which in the path-integral formulation is given by
\ba
Z&=&\int {\cal D}\psi^\dagger_s \,{\cal D}\psi_s \, e^{-S_E}\,,
\\
S_E&=&\int_0^\beta \!d\tau \int \! d^3{\bf x} \left[\psi^\dagger_s({\bf x}) \left(\partial_\tau-\frac{\nabla^2}{2m}-\mu\right)\psi_s({\bf x})+
   \frac{4 \pi a_s}{m} \psi^\dagger_\uparrow ({\bf x})\psi^\dagger_\downarrow ({\bf x})\psi_\downarrow ({\bf x})\psi_\uparrow ({\bf x})\right]\,,\nonumber
\ea
where $\beta,\mu$ are the inverse temperature and chemical potential of the system, respectively, and where we employ Einstein's sum convention for the spin index $s$. The 4-fermi interaction may be resolved by introducing a complex auxiliary field by inserting
\be
1=\int \!{\cal D}\zeta \,e^{\int_{\tau,{\bf x}} \frac{m}{4 \pi a_s} \zeta \zeta^*} 
\ee
inside the path integral, disregarding the irrelevant normalization. This converges for negative scattering length $a_s$. We subsequently shift the field by
$
\zeta\rightarrow \zeta- i \frac{4\pi a_s}{m} \psi_\downarrow \psi_\uparrow$  to find
\be
Z=\int {\cal D}\psi^\dagger_s \,{\cal D}\psi_s \,{\cal D}\zeta \,e^{-\int_{\tau,{\bf x}} \left[\psi^\dagger_s\left(\partial_\tau-\frac{\nabla^2}{2m}-\mu\right)\psi_s+ i \zeta^* \psi_\downarrow \psi_\uparrow-i\zeta \psi_\uparrow^\dagger \psi_\downarrow^\dagger-m \frac{\zeta \zeta^*}{4\pi a_s}\right]}\,.
\ee

\subsection{Thermodynamics}

All the fermionic bilinears may be collected in matrix form upon employing the Nambu-Gor'kov spinor $\Psi_s=\Bigg(\begin{array}{c}
  \psi_\uparrow\\
  \psi_\downarrow^\dagger
  \end{array}\Bigg)$, such that the effective Euclidean action becomes
\be
\label{seff1}
S_{E,{\rm eff}}=\int_{\tau,{\bf x}}\left[\Psi^\dagger\left(\partial_\tau-\sigma_z \frac{\nabla^2}{2m}-\sigma_z \mu + i\zeta^* \sigma_--i\zeta \sigma_+\right)\Psi -\frac{m \zeta \zeta^*}{4\pi a_s}\right]\,,
\ee
where $\sigma_\pm=\frac{1}{2}\left(\sigma_x\pm i \sigma_y\right)$ and $\sigma_i$ denote the Pauli matrices. Since the fermions are now quadratic, they may be integrated out, finding
\be
Z=\int {\cal D}\zeta  e^{\int_{\tau,{\bf x}} m \frac{\zeta \zeta^*}{4\pi a_s}+\ln {\rm det} \left[-G^{-1}(\zeta,\zeta^*)\right]}\,,
\ee
where $G^{-1}(\zeta,\zeta^*)=\partial_\tau-\sigma_z \frac{\nabla^2}{2m}-\sigma_z \mu + i\zeta^* \sigma_--i\zeta \sigma_+$.

So far, no approximations have been made. In order to perform the functional integral over $\zeta$, one can expand these fields around the global zero mode, e.g. $\zeta(\tau,{\bf x})=i \Delta+\zeta^\prime (\tau,{\bf x})$. Neglecting the contribution from the field fluctuations $\zeta^\prime$ corresponds to the R0 approximation
\cite{Romatschke:2019rjk} (or equivalently the leading large $N$ approximation in~\cite{veillette2007large}). Assuming $\Delta$ to be real, one finds
\be
Z_{R0}=\int_{-\infty}^\infty \! d\Delta \, e^{\beta V\left[\frac{m \Delta^2}{4\pi a_s}+\beta^{-1}\sum_{\omega_n}\int_{\bf k}\ln \det(- G^{-1}(\omega_n,{\bf k},\Delta))\right]}\,,
\ee
where the sum is over the fermionic Matsubara frequencies $\omega_n=(1+2n) \pi T$ with $n\in \mathbb{Z}$, $V$ is the volume of the system, and the inverse propagator in Fourier space is given by
\be
\label{ginv}
G^{-1}(\omega_n,{\bf k},\Delta)=i \omega_n+\sigma_z (\epsilon_{\bf k}-\mu)+\Delta \sigma_x\,,
\ee
where $\epsilon_{\bf k}\equiv \frac{{\bf k}^2}{2m}$.

In the large volume limit, the integral over $\Delta$ in $Z_{R0}$ is given exactly by the saddle point of the integral. The saddle point condition is
\be
\label{saddle}
\frac{m  \Delta}{4\pi a_s}+\beta^{-1}\sum_{\omega_n}\int \frac{d^3 {\bf k}}{(2\pi)^3}\frac{\Delta}{(\epsilon_{\bf k}-\mu)^2+\omega_n^2+\Delta^2}=0\,.
\ee
Besides the trivial solution $\Delta=0$, one can look for non-trivial solutions of this gap equation. In particular, we will consider the  zero temperature case.

For the case of zero temperature, the Matsubara sum turns into an integral, and the pressure becomes
\be
p=\frac{m \Delta^2}{4\pi a_s}+\int \frac{d\omega d^3k}{(2\pi)^4} \ln \left[(\epsilon_{\bf k}-\mu)^2+\omega^2+\Delta^2\right]\,.
\ee
The frequency integral is divergent, but being  field-theorists, we can employ dimensional regularization, which effectively means we are replacing $\ln \left[(\epsilon_{\bf k}-\mu)^2+\omega^2+\Delta^2\right]$ by $\ln \frac{\left[(\epsilon_{\bf k}-\mu)^2+\omega^2+\Delta^2\right]}{\omega^2}$, because $\int d\omega \ln \omega^2=0$ in dimensional regularization. In dimensional regularization, one thus finds
\be
\label{fmudelta}
p=\frac{m \Delta^2}{4\pi a_s}+\int \frac{d^3k}{(2\pi)^3} \sqrt{(\epsilon_{\bf k}-\mu)^2+\Delta^2}\,.
\ee
The integral over momenta is still divergent, but we will again employ dimensional regularization to extract the physically meaningful piece. Writing
\be
\label{esum}
\sqrt{(\epsilon_{\bf k}-\mu)^2+\Delta^2}=\sum_{n=0}^\infty \left(\begin{array}{c}
  \frac{1}{2}\\
n \end{array}\right) \frac{(-2 \mu \epsilon_{\bf k})^n}{\left(\epsilon_{\bf k}^2+\mu^2+\Delta^2\right)^{n-\frac{1}{2}}}\,,
\ee
all momentum integrals can be done in dimensional regularization, finding \cite{Nishida:2006eu}
\be
\label{mf}
\int \frac{d^{D}{\bf k}}{(2\pi)^D} \frac{\epsilon_{\bf k}^\alpha}{(\epsilon_{\bf k}^2+A^2)^{\frac{\beta}{2}}}=
\frac{\Gamma\left(\frac{D+2\alpha}{4}\right)\Gamma\left(\frac{2\beta-2\alpha-D}{4}\right)}{
  2 \Gamma\left(\frac{D}{2}\right)\Gamma\left(\frac{\beta}{2}\right)}
\left(\frac{m A}{2\pi}\right)^{\frac{D}{2}} A^{\alpha-\beta}\,.
\ee
Taking the limit $D\rightarrow 3$, it is curious to find that the resulting sum (\ref{esum}) can be evaluated in closed form, so that
\be
\int \frac{d^3k}{(2\pi)^3} \sqrt{(\epsilon_{\bf k}-\mu)^2+\Delta^2}=\frac{2}{5}\frac{\mu (2m \mu)^{\frac{3}{2}}}{3\pi^2} g\left(\frac{\mu}{\sqrt{\mu^2+\Delta^2}}\right)\,, 
\ee
where the function $g(y)=y^{-\frac{5}{2}}\left[(4 y^2-3)E\left(\frac{1+y}{2}\right)+\frac{3 +y-4 y^2}{2} K\left(\frac{1+y}{2}\right)\right]$, and $E,K$ are the complete elliptic integral of the first and second kind, respectively.  For the zero temperature case, in terms of $y=\frac{\mu}{\sqrt{\mu^2+\Delta^2}}$, the non-trivial solution to the gap equation thus has to fulfill
\be
\label{g2}
y^3 g^\prime(y) =\frac{15\pi}{8 \sqrt{2 m \mu}a_s}\,.
\ee
It is straightforward to check that the non-trivial solution to (\ref{g2}) has lower free energy than the trivial solution $\Delta=0$ for all $a_s<0$. For the unitary Fermi gas, $a_s\rightarrow -\infty$, and the solution to the gap equation requires $g^\prime(y)=0$ or
$
K\left(\frac{1+y}{2}\right)=2 E \left(\frac{1+y}{2}\right)$, for 
which a numerical solution gives $y\simeq 0.652$ or $\Delta\simeq 1.1622 \mu$. For small scattering lengths, the solution is close to $y\simeq 1$, and we can expand the left-hand-side of (\ref{g2}) near this point to find an analytic solution for the gap
\be
\label{Da}
\Delta=e^{-\frac{\pi}{\sqrt{8 m \mu a_s^2}}-2 +3 \ln 2}\times \mu \,,\quad  |a_s|\ll 1 \,,
\ee
which is within ten percent of the numerical solution of (\ref{g2}) for all $a_s<0$, and which gives $\Delta/\mu=e^{-2+3 \ln 2}\simeq 1.083$ for the cold unitary Fermi gas limit $a_s\rightarrow -\infty$. The presence of a non-vanishing gap indicates superfluidity, and the R0 approximation for $\Delta$ gives the right order of magnitude for the gap (see~\cite{gurarie2007resonantly} for a review on resonantly paired superfluids). To get an estimate of the accuracy of the approximation, we evaluate the pressure near unitarity,
\be
p=\frac{2}{5}\frac{\mu (2 m \mu)^{\frac{3}{2}}}{3\pi^2} \left.\frac{E\left(\frac{1+y}{2}\right)}{y^{\frac{3}{2}}}\right|_{y=(1+\frac{\Delta^2}{\mu^2})^{-\frac{1}{2}}}\simeq  2.2 \times p_{\rm free}\,,
\ee 
where $p_{\rm free}=\frac{2}{5}\frac{\mu (2 m \mu)^{\frac{3}{2}}}{3\pi^2}$ is the pressure of a non-interacting cold Fermi gas. Since it is known that the pressure at unitarity must be equal to $p=p_{\rm free}\xi^{-\frac{3}{2}}$ with constant Bertsch parameter $\xi$~\cite{Giorgini:2008zz}, we find $\xi_{R0}=y E^{-\frac{2}{3}}\left(\frac{1+y}{2}\right)\simeq 0.59$ for the R0 approximation, which should be compared to the established value $\xi\simeq 0.38$~\cite{Ku_2012,Endres:2012cw}.

From the pressure, one can define the number density of the Fermi gas near unitarity,
\be
\label{nuden}
\lim_{a_s\rightarrow - \infty}n(\mu)=\frac{\partial p}{\partial\mu}=\frac{(2m \mu)^{\frac{3}{2}}}{3\pi^2} \xi^{-\frac{3}{2}}\,,
\ee
which in turn is used to define the so-called Fermi momentum
\be
\label{kF}
k_F\equiv \left(3\pi^2 n(\mu)\right)^{\frac{1}{3}}\,.
\ee

\subsection{Transport Coefficient from Stress Tensor Correlators}

As stated above, Eq.~(\ref{tij}) implies that a gravitational wave passing through matter will create a local non-vanishing expectation value for the stress tensor, $\langle T^{ij}\rangle$. In order to calculate $\kappa$, however, it is not necessary to expose a system to an actual gravitational wave. The reason for this is that the coefficient of terms such as Eq.~(\ref{tij}) in the stress-tensor expectation value \textit{also appear in the stress-tensor correlation function in flat space}. So in order to calculate $\kappa$, we are considering the flat-space two-point correlation function $\langle T^{xy} T^{xy}\rangle$, which has been calculated before in a hydrodynamic gradient expansion. Note that the corresponding calculation is similar to the spectral function calculated in the large $N$ approximation in~\cite{Veillette_2008}.

From~\cite[Eq.~(2.124)]{Romatschke:2017ejr} one finds the real-time retarded correlator
\be
\label{corr}
C_R(\omega,k)\equiv\langle T^{xy} T^{xy}\rangle_R (\omega,k {\bf e}_z)=p-\frac{i \eta \omega+\frac{\kappa}{2} k^2 +\left(\frac{\kappa}{2}-\kappa^*\right)\omega^2}{1-i\tau_\pi \omega} + {\cal O}(\omega^4,k^4)\,,
\ee
where $p$ is the pressure, $\eta$ is the shear viscosity coefficient, $\kappa$ is the gravitational-wave to matter coupling sought in this work, and $\kappa^*$ is a second gravitational-wave to matter coupling coefficient relevant for \textit{bulk rather than shear stress}. We leave determination of $\kappa^*$ to future work.

Stress-tensor correlators are notoriously difficult to calculate from first principle in general, but the determination of $\kappa$ allows for a tremendous simplification. Namely, one notes that $\kappa$ (unlike say the viscosity coefficient $\eta$) couples to the wave-number $k$ in (\ref{corr}), so that it is sufficient to only calculate the zero-frequency correlator $C_R(\omega=0,k)$. This, in turn, means that it is not even necessary to perform an analytic continuation to real time, but instead it is sufficient to obtain $\kappa$ directly from the Euclidean correlation function evaluated at vanishing external Matsubara frequency,
\be
\label{kappa}
\kappa=\left.\frac{\partial^2}{\partial k^2} C_E(\omega_n=0,k)\right|_{k=0}\,,
\ee
cf.~\cite{Kovtun:2018dvd,Romatschke:2019gck}\footnote{Note that unlike for the case of bosonic theories, there is no additional term in (\ref{kappa}) because fermions do not allow for conformal coupling terms in the action. Also note that the Euclidean correlator $C_E(\omega_n)$ is \textit{minus} the retarded real-time correlator $C_R(\omega)$.}.
 
Having thus defined how to obtain $\kappa$ from flat-space correlation functions, we proceed in the next section to calculate $C_E(\omega_n,{\bf k})$ for the cold superfluid Fermi gas.

\subsection{Stress Tensor Correlators in the Cold Superfluid Fermi Gas}

The Euclidean stress-tensor two-point correlator for the Hamiltonian (\ref{hamiltonian}) is given in the path integral formulation as
\be
C_E(x-y) = Z^{-1} \int {\cal D}\psi^\dagger \psi e^{-S_E} T^{12}(x) T^{12}(y)\,,
\ee
where we have switched notation from $T^{xy}$ to $T^{12}$ and from $\left\{t,{\bf x}\right\}$ to four-coordinate $x$, and dropped the spin index for better readability. To obtain $T^{12}$, one can either start with the energy-momentum tensor for relativistic (4-component Dirac, see e.g. \cite{Kovtun:2018dvd}) fermions and perform the non-relativistic limit, or use the non-relativistic form derived in e.g.~\cite{Geracie:2016dpu}. In any case, one finds
\be
T^{12}(x)=\frac{1}{4m}\left[\partial_1 \psi^\dagger \partial_2 \psi+\partial_2 \psi^\dagger \partial_1 \psi
  -\partial_1\partial_2 \psi^\dagger \psi-\psi^\dagger \partial_1\partial_2 \psi\right]-\frac{i s}{4m} \partial_k \Sigma_k\,,
\ee
where $\Sigma_k=\psi^\dagger \sigma_l \epsilon_{kl1} \partial_2 \psi-\partial_2\psi^\dagger \sigma_l \epsilon_{kl1}\psi +\psi^\dagger \sigma_l \epsilon_{kl2} \partial_1\psi-\partial_1\psi^\dagger \sigma_l \epsilon_{kl2} \psi$ is a spin-current contribution. Here $\epsilon_{ijk}$ denotes the 3-dimensional totally anti-symmetric Levi-Civita symbol and $s=\frac{1}{2}$ for a spin $\frac{1}{2}$ fermion.
Disregarding the explicit spin-current contribution for now, the stress-tensor component may be rewritten in terms of the Nambu-Gor'kov spinors (cf. Eq.~(\ref{seff1})) as
\be
T^{12}(x)=\frac{1}{4m}\left[\partial_1 \Psi^\dagger \sigma_z \partial_2 \Psi + \partial_2 \Psi^\dagger \sigma_z \partial_1 \Psi
- \partial_1 \partial_2 \Psi^\dagger \sigma_z \Psi- \Psi^\dagger \sigma_z \partial_1 \partial_2 \Psi
  \right]-\frac{i s}{4m}\partial_k \Sigma_k\,.
\ee
Using the same auxiliary fields as in (\ref{seff1}), the fermionic part of the effective action is quadratic, so that the two-point stress-tensor correlator may be re-written in terms of two-point functions of the fundamental fermion field $\Psi$ because of Wick's theorem:
\ba
C_E(x-y)&=&\frac{-1}{2 m^2 Z}\int {\cal D}\Psi^\dagger {\cal D} \Psi {\cal D}\zeta  e^{-S_{E,{\rm eff}}} {\rm tr}\left[\partial_1^x \partial_2^{y} G(y-x) \sigma_z \partial_1^y \partial_2^x G(x-y) \sigma_z\right.\nonumber\\
  &&\left.+\partial_1^x \partial_1^{y} G(y-x) \sigma_z \partial_2^y \partial_2^x G(x-y) \sigma_z
  \right]+{\rm spin\ current}\,,
\ea
where ${\rm tr}$ refers to the trace over spinor indices and 
\be
\langle \Psi(x) \Psi^\dagger(y)\rangle=G(x-y)\,.
\ee

In the R0 approximation, the path integral over auxiliaries $\zeta,\zeta^*$ simplifies again because only the global zero mode is kept. As a consequence, the R0-propagator from Eq.~(\ref{ginv}) becomes
\be
G(\omega_n,{\bf k})=\frac{1}{(\epsilon_{\bf k}-\mu)^2+\omega_n^2+\Delta^2}\left(\begin{array}{cc}
  \epsilon_{\bf k}-\mu-i\omega_n & \Delta\\
  \Delta & -\epsilon_{\bf k}+\mu-i\omega_n
\end{array}\right)\,,
\ee
and the Fourier-transformed stress-tensor correlator $C_E(k_E,{\bf k})$ at zero temperature becomes
\be
C_E(k_E,{\bf k}=k {\bf e}_3)=-\int_{-\infty}^\infty d\Delta \frac{ e^{\beta V p}}{m^2Z} \int \frac{d^4 p}{(2\pi)^4}{\bf p}_1^2 {\bf p}_2^2 {\rm tr} \left[G(\omega,{\bf p})\sigma_z G(k_E+\omega,{\bf p+k}) \sigma_z\right]+{\rm s.\ c.}\,,
\ee
where  the ordinary integral over $\Delta$ will once again restrict the value of $\Delta$ to the solution of the saddle point condition, approximately given by (\ref{Da}).
Evaluating the spinor trace and restricting to vanishing external Matsubara frequency $k_E=0$, one finds
\be
C_E(0,k)=-\frac{2}{m^2} \int \frac{d^4 p}{(2\pi)^4} \frac{{\bf p}_1^2 {\bf p}_2^2 \left[(\epsilon_{\bf p}-\mu)(\epsilon_{\bf k+p}-\mu)-\omega^2-\Delta^2\right]}{\left[(\epsilon_{\bf p}-\mu)^2+\omega^2+\Delta^2\right]\left[(\epsilon_{\bf p+k}-\mu)^2+\omega^2+\Delta^2\right]}+{\rm s.\ c.} \,,
\ee
where the R0-partition function in the numerator and denominator has canceled.

The spin-current contribution can be obtained as follows: first note that given $T^{12}(x)\propto \partial_k \Sigma_k$, and because $\partial_{1,2}e^{i {\bf k}\cdot x}=0$ for ${\bf k}= k {\bf e}_3$, only $T^{12}(x)=-\frac{i s}{4m}\partial_3 \Sigma_3$ contributes to $\kappa$. Using Nambu Gor'kov spinors, $\Sigma_3$ may be written as
\be
\Sigma_3=\Psi \sigma_x (\partial_1-i\partial_2) \Psi+\Psi^\dagger \sigma_x (\partial_1+i \partial_2) \Psi^\dagger\,.
\ee
Since $\Sigma_3$ is linear in either $\partial_1$ or $\partial_2$, linear contributions of $\Sigma_3$ to $C_E$ vanish after angular integration. Hence, the only non-vanishing contribution to $\kappa$ from the spin-current is ${\rm s.\ c.}=-\left(\frac{s}{4m}\right)^2 \partial_3^x \partial_3^y \langle \Sigma_3(x) \Sigma_3(y)\rangle$, or in Fourier space
\be
   {\rm s.\ c.}=\frac{s^2 {\bf k}^2}{2 m^2}\int\frac{d^4 p}{(2\pi)^4}  {\bf p}_1^2{\rm tr} \left[\sigma_x G(\omega,{\bf p})\sigma_x  G(-\omega,-{\bf p})\right]\,.
\ee

Taking two derivatives with respect to $k$ and performing the angular averages gives an integral expression for $\kappa$ from (\ref{kappa}). Performing the integral over frequencies as well then leads to
\be
\kappa =\frac{2\Delta^2}{105m}\int \frac{d^3 p}{(2\pi)^3} \epsilon_p^2
\frac{(\epsilon_p-\mu)^2\left(21\mu-5\epsilon_p\right)+\Delta^2 \left(21\mu-25\epsilon_p\right)}{\left[(\epsilon_p-\mu)^2+\Delta^2\right]^{\frac{7}{2}}}
+\frac{2 s^2 \Delta^2}{3m}\int \frac{d^3 p}{(2\pi)^3} \frac{\epsilon_p}{\left[(\epsilon_p-\mu)^2+\Delta^2\right]^{\frac{3}{2}}}
\,,
  \ee
  and the remaining integral over momenta is convergent. Expanding the numerator in powers of $\epsilon_p$, the momentum integral can be solved using (\ref{mf}), and one finds
  \be
  \label{kappagen}
  \kappa=\frac{(2 m \mu)^{\frac{3}{2}}}{3\pi^2 m} \times \left.\frac{2 y E\left(\frac{1+y}{2}\right)+(1-y) K\left(\frac{1+y}{2}\right)}{8 y^\frac{3}{2}}\right|_{y=\frac{\mu}{\sqrt{\mu^2+\Delta^2}}}\left(-\frac{2}{3}+4s^2\right)\,,
  \ee
  and where again $E,K$ are the complete elliptic integral of the first and second kind, respectively.
  Close to unitarity, where the gap equation (\ref{g2}) implies $K\left(\frac{1+y}{2}\right)=2 E \left(\frac{1+y}{2}\right)$, this simplifies to
  \be
  \lim_{a_s\rightarrow-\infty}\kappa=\frac{(2 m \mu)^{\frac{3}{2}}}{3\pi^2 m} \frac{1}{\xi_{R0}^{\frac{3}{2}}}\left(s^2-\frac{1}{6}\right)\,,
  \ee
  with $\xi_{R0}\simeq 0.59$ the Bertsch parameter in the R0 approximation. Since the only non-trivial dependence of $\kappa$ on $a_s$ is through the Bertsch parameter, we predict that the correct value of the gravitational-wave matter coupling coefficient for the cold Fermi gas near unitarity is
  \be
  \lim_{a_s\rightarrow-\infty}\kappa=\frac{(2 m \mu)^{\frac{3}{2}}}{3\pi^2 m} \frac{1}{\xi^{\frac{3}{2}}}\left(s^2-\frac{1}{6}\right)\,.
  \ee
 Yet another way is to employ the number density (\ref{nuden}) to find the surprisingly simple result $\lim_{a_s\rightarrow-\infty}\kappa=\frac{n}{m}\left(s^2-\frac{1}{6}\right)$. For a spin $s=\frac{1}{2}$ fermion we thus have 
  \be
  \label{result}
  \lim_{a_s\rightarrow-\infty}\kappa=\frac{n}{12m}\,,
  \ee
  for the gravitational-wave matter coupling coefficient of the cold Fermi gas near unitarity. Eq.~(\ref{result}) is our main result.

  \section{Gravitational Wave Strain Rate}

  Since a potential application of the result (\ref{result}) is to detect gravitational waves, let us quickly review the form of a gravitational strain for an astrophysical system of interest, namely that of a binary black hole system.

  To get started, one writes the Einstein equations in matter
  \be
  R_{\mu\nu}-\frac{1}{2}g_{\mu\nu}R = 8 \pi G T_{\mu\nu}^{\rm source}\,,
  \ee
  where again $R_{\mu\nu},R$ are Ricci tensor and Ricci scalar, respectively, $g_{\mu\nu}$ is the metric field, and $T_{\mu\nu}^{\rm source}$
  is the stress tensor sourcing the gravitational wave, $G$ is Newton's gravitational constant and $\mu=\left\{0,1,2,3\right\}$. For small deviations from flat space, one linearizes the Einstein equations in the metric field $g_{\mu\nu}$ around the Minkowski metric. Since we will be particularly interested in the $g_{12}$ channel where the Minkowski metric is zero, we find
  \be
  \left[-\partial_0^2+\nabla^2\right] g_{12}(t,{\bf x})=-16 \pi T_{12}^{\rm source}(t,{\bf x})\,.
  \ee
  Now let's consider an astrophysical system that generates a non-trivial $T_{12}^{\rm source}(t,{\bf x})$ (we will give more details below). In this case, the strain rate $g_{12}$ can be calculated from the retarded Greens function of the operator $-\partial_0^2+\nabla^2$, and we find
  \be
  g_{12}(t,{\bf x})=4 G \int d^3y \frac{T_{12}^{\rm source}(t-|{\bf x}-{\bf y}|,{\bf y})}{|{\bf x}-{\bf y}|}\,.
  \ee
  In Fourier space, this becomes
  \be
  g_{12}(\omega,{\bf k})=4 G \int d^3y e^{-i {\bf k}\cdot {\bf y}} T_{12}^{\rm source}(\omega,{\bf y})\int d^3{\bf x}  \frac{e^{i\omega |{\bf x}|-i {\bf k}\cdot {\bf x}}}{|{\bf x}|}\,,
  \ee
  where one can recognize the Fourier-transform of a spherical wave, $\int_{\bf x}  \frac{e^{i\omega |{\bf x}|-i {\bf k}\cdot {\bf x}}}{|{\bf x}|}=\frac{4\pi}{k^2-\omega^2}$. One thus has
  \be
  g_{12}(\omega,{\bf k})=\frac{16 \pi G}{k^2-\omega^2} T_{12}^{\rm source}(\omega,{\bf k})\,.
  \ee

  For a model astrophysical system, consider two equal mass black holes orbiting each other on a circle with radius $R$. We will ignore GR effects for the motion of the black holes, which is a decent approximation for the inspiral phase. Without loss of generality, the locations of black hole one and two are then given by
\begin{equation}
\vec{x}^{(1)}=R \left(\begin{array}{c}\cos\Omega t\\\sin\Omega t\\0\end{array}\right)\text{ and }\vec{x}^{(2)}=-\vec{x}^{(1)}\text.
\end{equation}
Newton's equation of motion for the two-body system relate the orbital frequency $\Omega$ to the mass $M$ and radius $R$ through
    \be
    \label{nu}
    \Omega=\sqrt{\frac{GM}{4 R^3}}\,.
    \ee
    The stress-tensor component for this system is simply given by
    \be
    T^{\rm source}_{12}(t,{\bf x})=M \dot{x}^{(1)}\dot{y}^{(1)}\delta\left(\vec{x}-\vec{x}^{(1)}\right)+M \dot{x}^{(2)}\dot{y}^{(2)}\delta\left(\vec{x}-\vec{x}^{(2)}\right)\,,
    \ee
    where the dot indicates a time-derivative.
    If we consider the direction of the gravitational wave along the ${\bf k}=k {\bf e}_3$ direction as in the rest of this work, the Fourier transform of $T_{12}$ becomes particularly simple, and one finds
    \be
    \label{g12}
  g_{12}(\omega,k {\bf e}_3)=-\frac{2 \pi^2 r_s^2}{R} \frac{\left[\delta(\omega-2 \Omega)+\delta(\omega+2 \Omega)\right]}{k^2-\omega^2}  \,,
  \ee
  where (\ref{nu}) has been used and we have recognized the Schwarzschild radius of an individual black hole as $r_s=2 G M$. Note that (\ref{g12}) does not depend on Newton's gravitational constant (it is hidden in the Schwarzschild radius $r_s$ of the black hole), and that the decreasing strength of a spherical wave with distance from its center is encoded in the factor $\frac{1}{k^2-\omega^2}$.

  Of particular interest then is a potential experimental measurement of
  \be
  \label{lr}
  \langle T^{12}(\omega,{\bf k})\rangle=-2C_R(\omega,{\bf k})g_{12}(\omega,{\bf k}) +{\cal O}\left((g_{12})^2\right)\,,
  \ee
  where $g_{12}$ is the gravitational wave strain component, $C_R$ is defined in (\ref{corr}), and (\ref{lr}) is nothing but the linear-response formula connecting the one-point and two-point correlation function of the energy-stress tensor, cf. \cite[Eq.~(2.95)]{Romatschke:2017ejr}. For equilibrium configurations, $\langle T^{12}\rangle=0$ as can be quickly verified from kinetic theory. Therefore, a non-vanishing experimental determination of $\langle T^{12}\rangle$ in an equilibrium system provides a potential experimental handle on the gravitational wave strain $g_{12}$.

  However, thermodynamic fluctuations (aka ``noise'') will also lead to fluctuations in $\langle T^{12}\rangle$. To estimate the signal to noise ratio sufficient for detecting a gravitational wave using an experimental measurement of $\langle T^{12}\rangle$, we consider the ratio of gravitational wave stress (\ref{tij}) to thermodynamic pressure $p=p_{\rm free} \xi^{-\frac{3}{2}}$,
  \be
  \sigma\equiv \frac{\kappa \left[R^{\langle 12 \rangle}-2 R^{t \langle 12 \rangle t}\right]}{p}=\kappa \frac{k^2+\omega^2}{2 p}=\frac{5}{12 \xi} \frac{k^2+\omega^2}{2 k_F^2}\,.
  \ee

  To give some example numbers, for two solar-mass black holes orbiting each other at $R=10 r_s$ with $r_s=2 G M$ the Schwarzschild radius of a single black hole, one has $r_s\simeq 3$ km and $\frac{\Omega}{2\pi}\simeq 175$ Hz. The gravitational wave strain (\ref{g12}) will be peaked around $\omega=2 \Omega$ and  $k=\omega$, with a wavelength of $\frac{2\pi}{k}\simeq 1000$ km. For a unitary Fermi gas with $\frac{2\pi}{k_F}\sim 1$ $\mu{\rm m}$ \cite{zwierlein2005vortices}, we therefore estimate
  \be
  \label{estimate}
  \sigma \simeq 10^{-24}\,,
  \ee
  which is comparable to the required accuracy for detecting gravitational waves with interferometers, but unrealistic with current ultracold atom experiments.
  
  \section{Discussion}
\label{sec:discussion}
  
The result (\ref{result}) implies a spin-dependent value for the gravitational-wave to matter coupling $\kappa$ for superfluid Fermi gases near unitarity. For a spin $\frac{1}{2}$ fermion, it is useful to compare our finding with that of Shukla for a free Dirac fermion \cite{Shukla:2019shf}:
\be
\label{kappaR}
\kappa=\frac{1}{24\pi^2}\left[\mu_R \sqrt{\mu_R^2-m^2}-m^2 \ln \frac{\mu_R+\sqrt{\mu_R^2-m^2}}{m}\right]\,,
\ee
where $\mu_R=m+\mu$ is the relativistic chemical potential. In the non-relativistic limit, $\mu\ll m$, and (\ref{kappaR}) becomes $\kappa=\frac{n}{12m}$ in agreement with our result (\ref{result}) for a non-relativistic fermion near unitarity. As a consequence, for spin $\frac{1}{2}$ non-relativistic fermions, the result $\kappa=\frac{n}{12m}$ is universal both in the free and unitarity limit (but likely not in between, cf. \ref{kappagen}).

  The analytic result (\ref{result}) is amenable to independent verification by other theoretical methods. For instance, Monte Carlo and density functional theory methods that have proven successful in other applications such as those reported in~\cite{Bulgac:2017bho,Richie-Halford:2020ooz,Lu:2019nbg} may be used to test (\ref{result}), as well as extend it to the case of finite temperature and/or extend it to the case of positive scattering length. Another interesting extension would be to consider calculating $\kappa$ for polarized Fermi gases \cite{Bulgac:2020asx}, which can be treated in complete analogy to the method discussed here \cite{veillette2007large}. 

  Moreover, the result (\ref{result}) provides a testing ground for novel methods aiming at beating or ameliorating the so-called sign problem in fermionic systems, see e.g. \cite{Alexandru:2018ddf,Alexandru:2018ngw}.
  
  Last but not least, there is the prospect of using the information provided by (\ref{result}) as a means towards designing novel gravitational-wave detectors using Fermi gases near unitarity. A crude estimate of the required signal-to-noise ratio in (\ref{estimate}) for using experimental measurements of $\langle T^{12}\rangle$ to detect gravitational waves from a solar mass binary black hole merger is discouraging, but does not preclude the possibility of detection using more cleverly designed experimental techniques in the future.

  \section{Acknowledgments}

  This work was supported by the Department of Energy, DOE award No DE-SC0017905. We would like to thank A.~Bulgac, J.~Drut, M.~Forbes, L.~Radzihovsky, A.~Shukla and M. Veillete for useful discussions.

\bibliography{ufg}
\end{document}